\documentclass[%
reprint,
amsmath,amssymb,
aps,
pra,
floatfix,
]{revtex4-2}

\usepackage{graphicx}
\usepackage{dcolumn}
\usepackage{bm}
\usepackage{subfigure}
\usepackage{diagbox}

\begin{document}
	\preprint{APS/123-QED}
	
	\title{Heralded deterministic Knill-Laflamme-Milburn entanglement generation for solid-state emitters via waveguide-assisted photon scattering}
	
	\author{Fang-Fang Du$^{1*}$, Xin-Shan Du$^{1}$, Zhuo-Ya Bai$^{2}$, and Qiu-Lin Tan$^{1}$}
	
	\email[]{Duff@nuc.edu.cn}
	\email[]{tanqiulin@nuc.edu.cn}
	
	\affiliation{$^1$Key Laboratory of Micro/nano Devices and Systems, Ministry of Education, North University of China, Tai Yuan 030051, China}
	\affiliation{$^2$Beijing National Research Center for Information Science and Technology, Department of Electronic Engineering, Tsinghua University, Beijing 100084, China}
	
	\date{\today}
	
	\begin{abstract}
		The realization of quantum networks that exploit multiqubit entanglement opens avenues for transformative applications in the realm of quantum communication.
		In the paper, we present a set of heralded deterministic protocols designed for the generation of two-qubit, three-qubit, and $N$-qubit Knill-Laflamme-Milburn (KLM) states by the photon scattering property in one-dimensional waveguide-emitter system. In each protocol, the auxiliary single photon functions as a universal interface to bridge all stationary  qubits. Our proposed protocols allow for the conversion of irregular scattering incidents occasioned by nonideal coupling and frequency detuning into detectable events by triggering the detectors, which mean that our protocols for the generation of arbitrary KLM states with the predictive operational character and high fidelities. Owing to the significant breakthroughs in the integration of quantum emitters with nanophotonic waveguides, our protocolfs possess ideal features that position them as the promising candidate for deployment in long-range multiqubit quantum networks systems.
	\end{abstract}
	\maketitle
	
	\section{Introduction} \label{sec1}
	
	Quantum entanglement plays a central role in quantum information science and finds widespread applications in areas, including distributed quantum computing \cite{lim2005repeat,jiang2007distributed,su2024heralded}, quantum secure direct communication \cite{zhao2024quantum,ying2025passive,ying2024passive,10440135,qi2019implementation,zhang2017quantum,hu2016experimental,zhou2020device}, quantum secret sharing \cite{zhang2025device,zhang2025memory}, and quantum teleportation \cite{duan2022control,liu2025quantum,gong2024optimal,lv2024demonstration,wang2023verification}. In recent years, the generation of quantum entangled states has gained significant attention, with substantial progress in both theoretical research \cite{yan2021feasible,qin2018exponentially,jiang2023practically,zhou2020purification,guo2025polarization} and experimental implementation \cite{friis2019entanglement,zhang2020experimental,sciara2021scalable,hong2020experimental}.
	One such state, the Knill-Laflamme-Milburn (KLM) state \cite{lemr2010experimental,okamoto2011realization,guanzon2024saturating}, possesses remarkably robust entangled nature and further offers significant value for deepening the understanding and application of quantum entanglement phenomena \cite{erhard2020advances}, 
	thereby opening new avenues for future quantum information processing \cite{Litao1,highdimensional3,high-dimensional3,2023Xiu}. Hence,
	the exploration of generating and manipulating KLM states  has propelled advancements in related technological domains, establishing the foundation for broader research and applications in quantum optics.
	
	In recent years, the preparation and extension of photon KLM states have garnered significant attention, particularly in the context of quantum computing \cite{ukai2011demonstration,li2016hyperparallel,yoshikawa2007demonstration,rietsche2022quantum}.
	In 2007, Sandu Popescu \cite{popescu2007knill} proposed a KLM-based quantum computing scheme utilizing neutral atoms. The approach is notable for its lack of necessity for controlled interactions involving internal energy levels of the atoms, offering new perspectives on the realization of quantum computing.
	In 2018, Li \emph{et al.} \cite{li2018engineering} advanced the area by investigating the generation of various two-body KLM states for neutral atom systems. By combining the spontaneous emission of excited Rydberg states with the Rydberg blockade mechanism, they introduced a method for stabilizing KLM states, achievable from any initial state.
	In 2021, Zheng \emph{et al.} \cite{zheng2021fast} introduced a novel scheme leveraging dissipative processes to rapidly prepare stable KLM states between a pair of Rydberg atoms. The method significantly reduces the preparation time for steady states, marking an important technological breakthrough for the practical application of photon KLM states.
	In 2024, came from Liu \emph{et al.} \cite{liu2024generation} proposed  three-particle KLM state generation using inverse rotating interactions within a system comprising two frequency-tunable flux qubits and a coplanar waveguide resonator. The approach opens new possibilities for generating multi-particle KLM states, laying the groundwork for future advancements in quantum computing.
	Together, these studies are driving the preparation and application of photon KLM states toward greater stability, speed, and efficiency, thereby advancing progress in the field of quantum computing.
	
	The study addresses the issue of multi-partite quantum entanglement within the framework of quantum electrodynamics (QED). In the context, waveguide quantum electrodynamics (WG-QED) explores the interactions between propagating field modes, particularly those involving adjacent quantum emitter and one-dimensional (1D) waveguide. Research in WG-QED spans traditional physical platforms, such as photonic crystal waveguides \cite{yu2019two,douglas2015quantum,song2018photon,gonzalez2015subwavelength,song2025tunable}, optical fibers \cite{liedl2023collective,sorensen2016coherent,song2021optical,wang2019phase,cheng2017waveguide}, diamond-based waveguides \cite{sipahigil2016integrated,clevenson2015broadband}, superconducting transmission lines \cite{song2019microwave,yin2022non,sundaresan2019interacting,kannan2023demand}, and plasmonic nanowires \cite{yang2024non,akselrod2014probing}, as well as the investigation of emerging quantum emitters. These emitters can be realized through various means. 
	Recent studies have demonstrated that WG-QED not only provides a new platform for quantum information processing but also offers promising prospects for the realization of quantum networks \cite{kannan2020generating,chen2018entanglement,gajewski2021dissipation,zhang2019heralded,chen2017dissipation,mok2020long}. These developments have been instrumental in advancing the technological landscape of quantum communication, quantum computing, and related fields, further solidifying the pivotal role of WG-QED in the broader domain of quantum technologies \cite{holzinger2022control,sheremet2023waveguide,zhang2020subradiant,henriet2019critical,nie2023non}.
	
	The paper presents heralded deterministic protocol for the generation of two-qubit KLM state
	based on the imperfect scattering property of quantum emitters. In the approach, each quantum emitter is coupled to the corresponding 1D WG, and the degenerate ground states encode the qubits. The scattering errors, i.e., frequency detuning, weak coupling, and decay into non-ideal modes, can be converted into detectable photon polarization signals,
	consequently, the proposed KLM state generation scheme is highly predictable and can be directly extended to generate three-qubit and $N$-qubit KLM states.
	
	The structure of the paper is organized as follows. First, we propose a scheme for implementing a Z gate using an emitter restricted within a 1D WG in Sec. \ref{sec2}. Next, leveraging the interaction between a single photon and a quantum emitter, we introduce three protocols for generating the two-qubit, three-qubit, and $N$-qubit KLM states in sequence  in Sec. \ref{sec3}. Finally, we analyze the success probability of the scheme in practical systems and conclude the paper in Sec. \ref{sec4}.
	
	\section{The scattering of photon off single emitter}\label{sec2}
	
	\begin{figure}[htp]
		\centering
		\subfigure[]{
			\begin{minipage}{1\linewidth}
				\centering
				\includegraphics[width=0.8\linewidth]{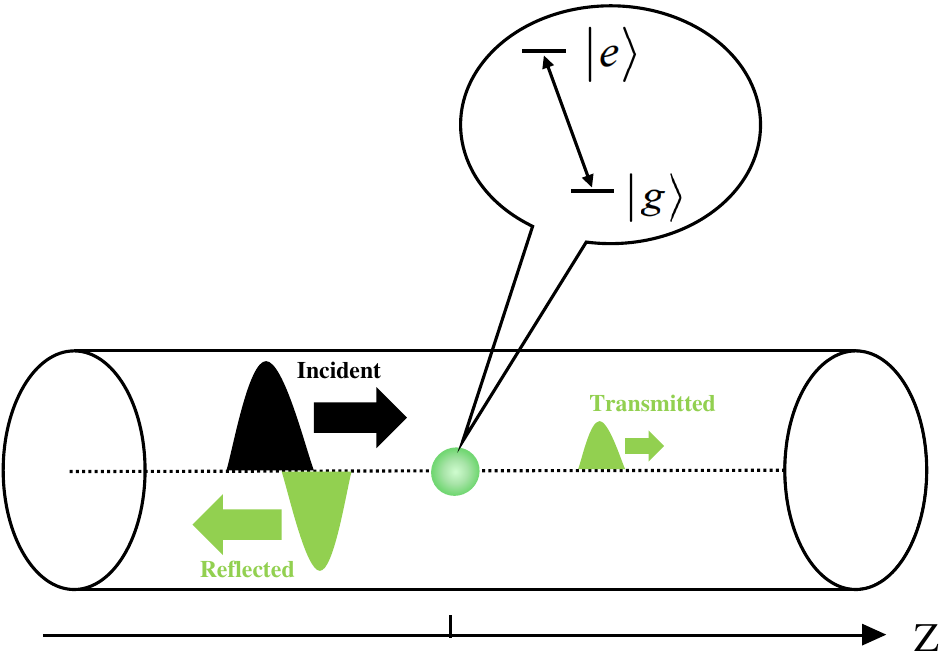}
				\label{fig1(a)}
		\end{minipage}}
		\subfigure[]{
			\begin{minipage}{1\linewidth}
				\centering
				\includegraphics[width=0.8\linewidth]{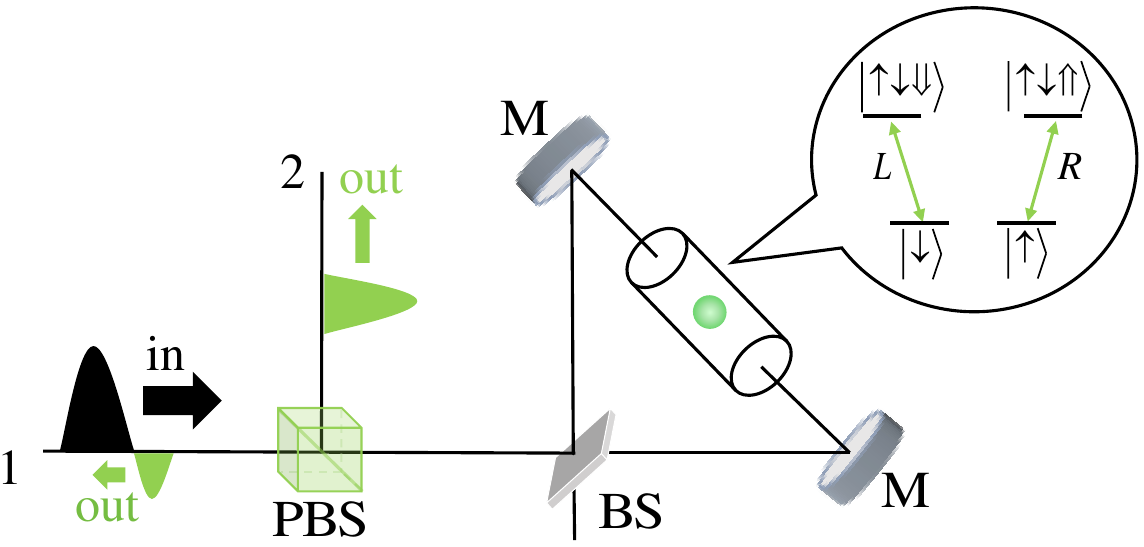}
				\label{fig1(b)}
		\end{minipage}}
		\caption{(a) The diagram illustrates a two-level emitter (depicted as a green dot) coupled to a 1D WG (represented by the cylinder). An incoming photon (depicted in black) from the left undergoes scattering with the emitter, producing both reflected and transmitted components.
			(b) A heralded configuration is presented for the implementation of a Z gate using an emitter restricted within a 1D WG. PBS is a polarized beam splitter, which reflects $|V\rangle$-polarized (transmits $|H\rangle$-polarized)
			photon. BS is a balanced beam splitter. M is a reflected mirror. The inset provides details on the optical transitions and level structure of the quantum emitter.} \label{fig1}
	\end{figure}
	
	The interaction between individual photons and isolated emitters is crucial for achieving quantum entanglement. In Fig. \ref{fig1}(a), the two-level emitter is intricately coupled to a 1D WG through an electromagnetic mode characterized by an intensity parameter, with the system state represented as $|V\rangle$. The emitter is constituted by an excited state $ |e\rangle $ and a ground state $ |g\rangle $, featuring a transition frequency denoted as $ \omega_{a} $. In a 1D quantum system, the Hamiltonian can be expressed as Eq. (\ref{eq1}) when $ \omega_{a} $ deviates from the cutoff frequency defined by the dispersion relation \cite{shen2005coherent}.
	\begin{eqnarray} \label{eq1}   
		H&=&(\omega_{a}-\frac{i\varGamma^{'}_{e}}{2})\sigma_{ee}\nonumber\\
		&&+V\int dz\delta(z)\{{\sigma_{{eg}}[c_{_{R}}(z)+c_{_{L}}(Z)]+H.c.}\}\nonumber\\
		&&+iv_{g}\int dz[c^{\dagger}_{_{L}}(z)\frac{\partial c_{_{L}}(z)}{\partial z}-c^{\dagger}_{_{R}}(z)\frac{\partial c_{_{R}}(z)}{\partial z}].
	\end{eqnarray}
	Here, $\nu_{g}$ refers to the group velocity, while $\varGamma^{'}_{e}$ is the decay rate from the excited state to free space. The annihilation operators for the left and right propagating fields are denoted by $c_{_{L}}$ and $c_{_{R}}$, respectively.
	
	We assume that a photon with energy $E_{k}$ is incident from the left, resulting in
	\begin{eqnarray} \label{eq2}
		|E\rangle _{k}&=&\int dx[\phi_{_{L}}(x)c^{\dagger}_{_{L}}(x)
		+\phi_{_{R}}(x)c^{\dagger}_{_{R}}(x)]|g,vac\rangle\nonumber\\
		&&+c_{e}|e,vac\rangle
		,
	\end{eqnarray}
	where $c^{\dagger}_{_{L}}(x)$ and $c^{\dagger}_{_{R}}(x)$ are the bosonic operators for the photon, $ |vac\rangle $ denotes the vacuum state of the photon, and $c_{e}$ represents the probability amplitude. The fields $\phi_{_{L,R}}(x)$ can be formulated as
	\begin{eqnarray} \label{eq3}   
		&&\phi_{_{L}}(x)=re^{-ike}\theta(-x),\nonumber\\
		&&\phi_{_{R}}(x)=te^{ikx}\theta(x)+e^{ikx}\theta(-x).
	\end{eqnarray}
	Here, $\theta(x)$ represents the Heaviside step function. By solving the Schr\"{o}dinger equation, the following results can be obtained \cite{shen2005coherent}
	\begin{eqnarray} \label{eq4}   
		r=\frac{-1}{1-2i\Delta/\gamma_{_{1D}}+\gamma^{'}/\gamma_{_{1D}}},\;\;\;
		t=r+1,
	\end{eqnarray}
	where $\gamma_{_{1D}}=4\pi g^{2}/c$ represents the decay rate, and $\Delta$ denotes the detuning. When $\Delta=0$ (resonance) and the Purcell factor $\gamma_{_{1D}}/\gamma^{'}\gg1$, the transmission and reflection coefficients are approximately $t\approx0$ and $r\approx-1$, respectively. In contrast, when $\Delta\neq0$, the incident photon is unaffected by the emitter.
	
	In the system depicted in Fig. \ref{fig1}(b), we examine an emitter that consists of two excited states and two ground states. The ground states are labeled as $|\uparrow\rangle=|g_{+}\rangle$ and $|\downarrow\rangle=|g_{-}\rangle$, while the excited states are denoted as $|\uparrow\downarrow\Uparrow\rangle=|e_{+}\rangle$ and $|\uparrow\downarrow\Downarrow\rangle=|e_{-}\rangle$ for simplicity. When a photon is injected from the left into the system, it can be in either a horizontal polarization state $|H\rangle$ or a vertical polarization state $|V\rangle$. Based on the previously established scattering properties, the scattering process can be described as follows
	\begin{eqnarray} \label{eq5}   
		&&|g_{\pm}\rangle|\psi\rangle|H^{1}\rangle\rightarrow|g_{\pm}\rangle|\psi_{t}\rangle|H^{1}\rangle\pm|g_{\pm}\rangle|\psi_{r}\rangle|V^{2}\rangle,\nonumber\\
		&&|g_{\pm}\rangle|\psi\rangle|V^{2}\rangle\rightarrow|g_{\pm}\rangle|\psi_{t}\rangle|V^{2}\rangle\pm|g_{\pm}\rangle|\psi_{r}\rangle|H^{1}\rangle.
	\end{eqnarray}
	where $|\psi_{t}\rangle=t|\psi\rangle$ and $|\psi_{r}\rangle=r|\psi\rangle$.
	
	Utilizing specific optical components and integrating the previously discussed principles, a heralded Z-gate can be fabricated. The emitter is initially set in one of the two states $|g_{\pm}\rangle $, and a photon, characterized by the polarization state $|H\rangle$ (for port 1) or $|V\rangle$ (for port 2) is injected and the spatial state $|\psi\rangle$. In the heralded Z gate, the interaction between the emitter and the photon is as follows
	\begin{eqnarray} \label{eq6}   
		&&|g_{\pm}\rangle|\psi\rangle|H^{1}\rangle\rightarrow\pm|g_{\pm}\rangle|\psi_{r}\rangle|V^{2}\rangle,\nonumber\\
		&&|g_{\pm}\rangle|\psi\rangle|V^{2}\rangle\rightarrow\pm|g_{\pm}\rangle|\psi_{r}\rangle|H^{1}\rangle.
	\end{eqnarray}
	Specifically, in the ideal scenario where $|\psi_{r}\rangle=-|\psi\rangle$, and disregarding the spatial states of the photons, Eq. (\ref{eq6}) can be reformulated as
	\begin{eqnarray} \label{eq7}   
		|g_{\pm}\rangle|H^{1}\rangle\rightarrow\mp|g_{\pm}\rangle|V^{2}\rangle,\;\;\;\;
		|g_{\pm}\rangle|V^{2}\rangle\rightarrow\mp|g_{\pm}\rangle|H^{1}\rangle.
	\end{eqnarray}
	
	If the emitter coupled with the 1D
	WG is set in the superposition state  $|\pm\rangle=\frac{1}{\sqrt{2}}(\vert g_{+}\rangle\pm\vert g_{-}\rangle)$, the evolution induced by scattering follows
	\begin{eqnarray} \label{eq}
		|\pm\rangle \vert H^{1}\rangle\rightarrow  r\vert \mp \rangle\vert V^{2}\rangle,\;\;\;\;
		|\pm\rangle \vert V^{2}\rangle\rightarrow  r\vert \mp \rangle\vert H^{1}\rangle.
	\end{eqnarray}
	
	\section{the generation of heralded deterministic KLM states for solid-state emitters}\label{sec3}
	
	\subsection{ The generation of heralded deterministic two-qubit KLM state }\label{sec3.1}
	
\begin{figure}[htpb]
		\centering
		\begin{center}
			\centering
			\includegraphics[width=1\linewidth]{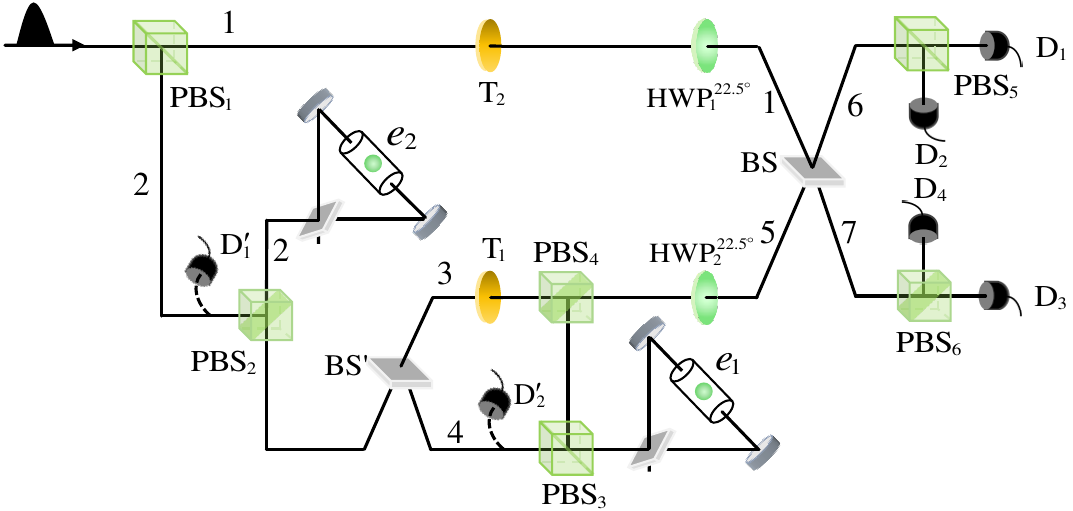}
			\caption{Schematic representation for generating the two-qubit KLM state. The BS$'$ is used to perform the conversion between the upper $(u)$ and lower $(l)$ spatial modes, i.e., $|m_{u}\rangle\rightarrow(|m_{l}\rangle-|m_{u}\rangle)/\sqrt{2}$ and $|m_{l}\rangle\rightarrow(|m_{u}\rangle+|m_{l}\rangle)/\sqrt{2}$. Another BS performs the transformation
				$|m_{u}\rangle\rightarrow(|m_{u}\rangle+|m_{l}\rangle)/\sqrt{2}$ and $|m_{l}\rangle\rightarrow(|m_{u}\rangle-|m_{l}\rangle)/\sqrt{2}$.
				HWP$_{i}^{22.5^{\circ}}(i=1, 2)$ refers to a half-wave plate rotated  to  $\theta=22.5^{\circ}$. T$_{j} (j=1,2)$ is a wave plate with a specific transmission coefficient $r^{j}$. D$ _{q} ( q=1, 2, 3, 4$) is a single-photon detector.
			}\label{fig2}
		\end{center}
	\end{figure}

	The configuration for the two-emitter KLM state is outlined  in Fig. \ref{fig2}. Initially, the  auxiliary single-photon state is prepared in the state $|\psi^{p}\rangle = \frac{1}{\sqrt{3}}|H\rangle+\frac{\sqrt{2}}{\sqrt{3}}|V\rangle $. Alternatively, the single photon in the  $|H\rangle$-polarization state may be introduced, then it undergoes a operation via a $27.4^{\circ}$ half-wave plate (HWP),
	where the matrices of the HWP$^{\theta}$ rotated to $\theta$ angle in the basis $\{|H\rangle$, $|V\rangle\}$ is  given by
	\begin{align} \label{eq0}
		\text{HWP}^{\theta}=&{\left[ \begin{array}{cccc}
				\cos 2\theta & \sin 2\theta \\
				\sin 2\theta & -\cos 2\theta
			\end{array}
			\right ]}.
	\end{align}
	Besides, each of the two emitters coupled with the 1D WG is set in the state $|+\rangle_{k}=\frac{1}{\sqrt{2}}(|g_{+}\rangle+|g_{-}\rangle)_{k}$ $(k=e_{1},e_{2})$. Thus, the overall system state is expressed as $|\Lambda\rangle_{0}=(\frac{1}{\sqrt{3}}|H\rangle+\frac{\sqrt{2}}{\sqrt{3}}|V\rangle)|+\rangle_{e_{1}}|+\rangle_{e_{2}} $.
	
	Firstly, after the single photon enters the first PBS$_{1}$, the photon in the $|H\rangle$-polarization state is transmitted, while the photon in the $|V\rangle$-polarization state is reflected and directed to the second PBS$_{2}$. Following then, the overall system state evolves from $|\Lambda\rangle_{0}$ to $|\Lambda\rangle_{1}$
	\begin{eqnarray} \label{eq8}
		|\Lambda\rangle_{1} = \frac{1}{\sqrt{3}}|H^{1}\rangle|+\rangle_{e_{1}}|+\rangle_{e_{2}}+\frac{\sqrt{2}}{\sqrt{3}}|V^{2}\rangle|+\rangle_{e_{1}}|+\rangle_{e_{2}}.
	\end{eqnarray}
	
	\begin{table*}
		\centering
		\caption{The measurement results of the ancillary photon state, and the corresponding two-emitter state. The feedforward	operations are given by $I^{e_{1},e_{2}}=|+\rangle\langle+|+|-\rangle\langle-|$, $\sigma_{z}^{e_{1},e_{2}}=|+\rangle\langle+|-|-\rangle\langle-|$.} \label{Table1}
		\setlength{\tabcolsep}{11pt}
		\renewcommand\arraystretch{2}
		\begin{tabular} {c c c c}
			\hline
			D$_{q}$ & Ancillary photon state & Two-emitter state  & Feedforward operations \\
			\hline
			D$_{1}$ & $|H^{6}\rangle$ &  $\frac{1}{\sqrt{3}}(|+\rangle_{e_{1}}|+\rangle_{e_{2}}+|+\rangle_{e_{1}}|-\rangle_{e_{2}}+|-\rangle_{e_{1}}|-\rangle_{e_{2}})$ & $I^{e_{1}}\otimes I^{e_{2}}$ \\
			D$_{2}$ & $|V^{6}\rangle$ &  $\frac{1}{\sqrt{3}}(|+\rangle_{e_{1}}|+\rangle_{e_{2}}+|+\rangle_{e_{1}}|-\rangle_{e_{2}}-|-\rangle_{e_{1}}|-\rangle_{e_{2}})$ & $\sigma_{z}^{e_{1}}\otimes I^{e_{2}}$ \\
			D$_{3}$ & $|H^{7}\rangle$ &  $\frac{1}{\sqrt{3}}(|+\rangle_{e_{1}}|+\rangle_{e_{2}}-|+\rangle_{e_{1}}|-\rangle_{e_{2}}+|-\rangle_{e_{1}}|-\rangle_{e_{2}})$ & $\sigma_{z}^{e_{1}}\otimes\sigma_{z}^{e_{2}}$ \\
			D$_{4}$ & $|V^{7}\rangle$ &  $\frac{1}{\sqrt{3}}(|+\rangle_{e_{1}}|+\rangle_{e_{2}}-|+\rangle_{e_{1}}|-\rangle_{e_{2}}-|-\rangle_{e_{1}}|-\rangle_{e_{2}})$ & $I^{e_{1}}\otimes\sigma_{z}^{e_{2}}$ \\
			\hline
		\end{tabular}
	\end{table*}
	
	Secondly, the photon in spatial mode 2 is reflected by PBS$_{2}$ and interacts with the emitter $e_{2}$ in 1D WG. Following the interaction, the photon's state is transformed to $|H\rangle$ and it passes sequentially through PBS$_{2}$ again and BS$'$. As a result of these processes, the following outcome is obtained without the response of the detector D$_{1}'$
	\begin{eqnarray} \label{eq9}
		|\Lambda\rangle_{2} &=&
		\frac{1}{\sqrt{3}}(|H^{1}\rangle|+\rangle_{e_{1}}|+\rangle_{e_{2}}+r|H^{3}\rangle|+\rangle_{e_{1}}|-\rangle_{e_{2}}\nonumber\\
		&&+r|H^{4}\rangle|+\rangle_{e_{1}}|-\rangle_{e_{2}}),
	\end{eqnarray}
	where $|-\rangle = \dfrac{1}{\sqrt{2}}(|g_{+}\rangle-|g_{-}\rangle)$.
	
	Thirdly, the photon in spatial mode 4 interacts with the emitter $e_{1}$ through PBS$_{3}$. If the D$_{2}'$ is not triggered, the photon state is altered to $|V\rangle$, and it is reflected back to PBS$_{4}$ by PBS$_{3}$. Meanwhile, the photon in spatial mode 1(3) passes through T$_{2}$ (T$_{1}$ and PBS$_{4}$), where T$_{j} (j=1,2)$ customizes the transmission coefficient $r^{j}$ for the photon. Subsequently, the photon in spatial mode 1(3) undergoes a Hadamard operation via HWP$ _{1}^{22.5^{\circ}}$ (HWP$ _{2}^{22.5^{\circ}} $), resulting in
	\begin{eqnarray} \label{eq10}
		|\Lambda\rangle_{3} &=&
		\frac{r^{2}}{\sqrt{6}}[(|H^{1}\rangle+|V^{1}\rangle)|+\rangle_{e_{1}}|+\rangle_{e_{2}}\nonumber\\
		&&+(|H^{5}\rangle+|V^{5}\rangle)|+\rangle_{e_{1}}|-\rangle_{e_{2}}\nonumber\\
		&&+(|H^{5}\rangle-|V^{5}\rangle)|-\rangle_{e_{1}}|-\rangle_{e_{2}}].
	\end{eqnarray}
	
	Fourthly, the photon in spatial modes 1 and 5 interferes at BS. As a result, the state of the system is collapsed to $|\Lambda\rangle_{4}$
	\begin{eqnarray} \label{eq11}
		|\Lambda\rangle_{4} &=&
		\dfrac{r^{2}}{2\sqrt{3}}[|H^{6}\rangle(|+\rangle_{e_{1}}|+\rangle_{e_{2}}+|+\rangle_{e_{1}}|-\rangle_{e_{2}}+|-\rangle_{e_{1}}|-\rangle_{e_{2}})\nonumber\\
		&&+|H^{7}\rangle(|+\rangle_{e_{1}}|+\rangle_{e_{2}}-|+\rangle_{e_{1}}|-\rangle_{e_{2}}-|-\rangle_{e_{1}}|-\rangle_{e_{2}})\nonumber\\
		&&+|V^{6}\rangle(|+\rangle_{e_{1}}|+\rangle_{e_{2}}+|+\rangle_{e_{1}}|-\rangle_{e_{2}}-|-\rangle_{e_{1}}|-\rangle_{e_{2}})\nonumber\\
		&&+|V^{7}\rangle(|+\rangle_{e_{1}}|+\rangle_{e_{2}}-|+\rangle_{e_{1}}|-\rangle_{e_{2}}+|-\rangle_{e_{1}}|-\rangle_{e_{2}})].\nonumber\\
	\end{eqnarray}
	
	Finally, after the photon traverses PBS$_{5}$ and PBS$_{6}$, it is detected by the detector D$_{q}$ $(q=1,2,3,4)$ in the $\{|H\rangle $, $ |V\rangle\}$ basis. Upon detection by D$_{1}$,  the state of two emitters is collapsed into the two-qubit KLM state
	\begin{eqnarray} \label{eq12}
		|\text{KLM}\rangle_{2} &=&\frac{r^{2}}{\sqrt{3}}(|+\rangle_{e_{1}}|+\rangle_{e_{2}}+|+\rangle_{e_{1}}|-\rangle_{e_{2}}+|-\rangle_{e_{1}}|-\rangle_{e_{2}}).\nonumber\\
	\end{eqnarray}
	For the response of the other detection D$_{q}$ $(q=2,3,4)$, the two-qubit KLM state in Eq. (\ref{eq12}) can be achieved by applying classical feedforward operations to the output quantum state,
	as illustrated in Table \ref{Table1}, based on the measurement results of the auxiliary photon.
	Therefore, the success probability  of generating the two-qubit KLM state is $p_{2}=|r^{2}|^{2}$. 
	
	\subsection{ The generation of heralded deterministic three-qubit KLM state}\label{sec3.2}
	\begin{figure*}
		\centering
		\begin{center}
			\centering
			\includegraphics[width=0.7\linewidth]{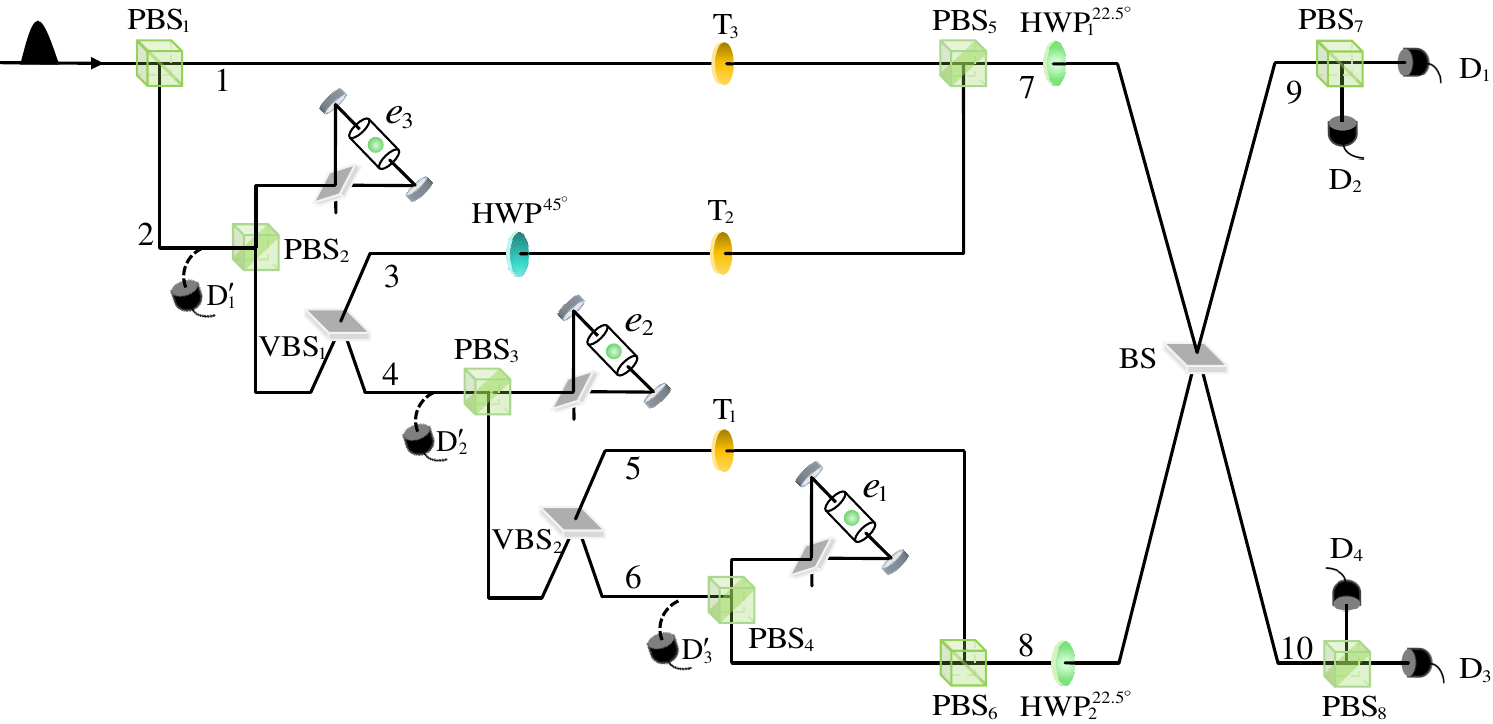}
			\caption{Schematic representation for generating the three-qubit KLM state.  VBS$_{k}(k=1, 2)$ represents unbalanced beam splitter, that is, $|m_{l}\rangle\rightarrow\frac{1}{\sqrt{4-k}}|m_{u}\rangle+\frac{\sqrt{3-k}}{\sqrt{4-k}}|m_{l}\rangle$, and $|m_{u}\rangle\rightarrow\frac{1}{\sqrt{4-k}}|m_{l}\rangle-\frac{\sqrt{3-k}}{\sqrt{4-k}}|m_{u}\rangle$.}\label{fig3}
		\end{center}
	\end{figure*}
	As illustrated in Fig. \ref{fig3}, the configuration for the three-emitter KLM state begins with the preparation of the auxiliary single-photon state, i.e., $|\psi^{p}\rangle = \frac{1}{2}|H\rangle+\frac{\sqrt{3}}{2}|V\rangle$, where the single photon in the $|H\rangle$-polarization state is subjected to the operation using the HWP$^{30^{\circ}}$. Meanwhile, each of the three emitters coupled with the 1D WG is prepared in the state $|+\rangle_{k}$ $(k=e_{1},e_{2},e_{3})$. As a result, the overall state of the system is
	$ |\Phi\rangle_{0}=(\frac{1}{2}|H\rangle+\frac{\sqrt{3}}{2}|V\rangle)\otimes|+\rangle_{e_{1}}|+\rangle_{e_{2}}|+\rangle_{e_{3}}$.

	Firstly, after the photon passes through PBS$_{1}$, the photon in spatial mode 2 is reflected by PBS$_{2}$ and then interacts with emitter $e_{3}$ in 1D WG. If the detector D$_{1}'$ in Fig. \ref{fig3} is not triggered, the state of the photon becomes $|H\rangle$, and it subsequently passes through PBS$_{2}$ and VBS$_{1}$. Through the sequence of steps, resulting in
	\begin{eqnarray} \label{eq13}
		|\Phi\rangle_{1} &=& \frac{1}{2}|H^{1}\rangle|+\rangle_{e_{1}}|+\rangle_{e_{2}}|+\rangle_{e_{3}}+\frac{r}{2}|H^{3}\rangle|+\rangle_{e_{1}}|+\rangle_{e_{2}}|-\rangle_{e_{3}}\nonumber\\
		&&+\frac{r}{\sqrt{2}}|H^{4}\rangle|+\rangle_{e_{1}}|+\rangle_{e_{2}}|-\rangle_{e_{3}}.
	\end{eqnarray}
	
	Secondly, the photon in spatial mode 4 passes through PBS$_{3}$ and interacts with emitter $e_{2}$ in 1D WG. The state  of the photon becomes $|V\rangle$ without the response of the detector D$_{2}'$ in Fig. \ref{fig3}, and the photon is reflected by PBS$_{3}$ towards VBS$_{2}$, leading to
	\begin{eqnarray} \label{eq14}
		|\Phi_{2}\rangle &=& \frac{1}{2}[|H^{1}\rangle|+\rangle_{e_{1}}|+\rangle_{e_{2}}|+\rangle_{e_{3}}+r|H^{3}\rangle|+\rangle_{e_{1}}|+\rangle_{e_{2}}|-\rangle_{e_{3}}\nonumber\\&&+r^{2}(|V^{5}\rangle|+\rangle_{e_{1}}|-\rangle_{e_{2}}|-\rangle_{e_{3}}
		+|V^{6}\rangle|+\rangle_{e_{1}}|-\rangle_{e_{2}}|-\rangle_{e_{3}})].\nonumber\\
	\end{eqnarray}
	
	Thirdly, the photon in spatial mode 1 passes through T$_{3}$, while the photon in spatial mode 3 passes through HWP$^{45^{\circ}}$ (shown in Eq. (\ref{eq0})) and T$_{2}$. The two spatial modes 1 and 3 are combined at spatial mode 7 by the PBS$_{5}$. Meanwhile, the photon in spatial mode 5 passes through T$_{1}$, while the photon in spatial mode 6 interacts with the emitter $e_{1}$ in 1D WG after being reflected by PBS$_{4}$. Then the photon in two spatial modes is combined at PBS$_{6}$ without the response of the detector D$_{3}'$. Subsequently, the photon in spatial modes 7 (8) undergoes the HWP$_{1}^{22.5^{\circ}}$ (HWP$ _{2}^{22.5^{\circ}}$). As a result, the following outcome is obtained
	\begin{eqnarray} \label{eq15}
		|\Phi\rangle_{3} &=& \frac{r^{3}}{2\sqrt{2}}[|H^{7}\rangle(|+\rangle_{e_{1}}|+\rangle_{e_{2}}|+\rangle_{e_{3}}+|+\rangle_{e_{1}}|+\rangle_{e_{2}}|-\rangle_{e_{3}})\nonumber\\
		&&+|V^{7}\rangle(|+\rangle_{e_{1}}|+\rangle_{e_{2}}|+\rangle_{e_{3}}-|+\rangle_{e_{1}}|+\rangle_{e_{2}}|-\rangle_{e_{3}})\nonumber\\
		&&+|H^{8}\rangle(|+\rangle_{e_{1}}|-\rangle_{e_{2}}|-\rangle_{e_{3}}+|-\rangle_{e_{1}}|-\rangle_{e_{2}}|-\rangle_{e_{3}})\nonumber\\
		&&+|V^{8}\rangle(-|+\rangle_{e_{1}}|-\rangle_{e_{2}}|-\rangle_{e_{3}}+|-\rangle_{e_{1}}|-\rangle_{e_{2}}|-\rangle_{e_{3}})].\nonumber\\
	\end{eqnarray}
	
	Fourthly, the photon of spatial modes 7 and 8 interferes at BS, resulting in
	\begin{eqnarray} \label{eq16}
		|\Phi\rangle_{4} &=& \dfrac{r^{3}}{4}[|H^{9}\rangle(|+\rangle_{e_{1}}|+\rangle_{e_{2}}|+\rangle_{e_{3}}+|+\rangle_{e_{1}}|+\rangle_{e_{2}}|-\rangle_{e_{3}}\nonumber\\
		&&+|+\rangle_{e_{1}}|-\rangle_{e_{2}}|-\rangle_{e_{3}}+|-\rangle_{e_{1}}|-\rangle_{e_{2}}|-\rangle_{e_{3}})\nonumber\\
		&&+|H^{10}\rangle(|+\rangle_{e_{1}}|+\rangle_{e_{2}}|+\rangle_{e_{3}}+|+\rangle_{e_{1}}|+\rangle_{e_{2}}|-\rangle_{e_{3}}\nonumber\\
		&&-|+\rangle_{e_{1}}|-\rangle_{e_{2}}|-\rangle_{e_{3}}-|-\rangle_{e_{1}}|-\rangle_{e_{2}}|-\rangle_{e_{3}})\nonumber\\
		&&+|V^{9}\rangle(|+\rangle_{e_{1}}|+\rangle_{e_{2}}|+\rangle_{e_{3}}-|+\rangle_{e_{1}}|+\rangle_{e_{2}}|-\rangle_{e_{3}}\nonumber\\
		&&-|+\rangle_{e_{1}}|-\rangle_{e_{2}}|-\rangle_{e_{3}}+|-\rangle_{e_{1}}|-\rangle_{e_{2}}|-\rangle_{e_{3}})\nonumber\\
		&&+|V^{10}\rangle(|+\rangle_{e_{1}}|+\rangle_{e_{2}}|+\rangle_{e_{3}}-|+\rangle_{e_{1}}|+\rangle_{e_{2}}|-\rangle_{e_{3}}\nonumber\\
		&&+|+\rangle_{e_{1}}|-\rangle_{e_{2}}|-\rangle_{e_{3}}-|-\rangle_{e_{1}}|-\rangle_{e_{2}}|-\rangle_{e_{3}})].
	\end{eqnarray}
	
	Finally, after the photon traverses PBS$_{7}$ and PBS$_{8}$, it is detected by the single-photon detector D$_{q}$ $(q=1,2,3,4)$ in the $\{|H\rangle $, $ |V\rangle\}$ basis. Upon detection by D$_{1}$, the state of three emitters is collapsed into the three-qubit KLM state
	\begin{eqnarray} \label{eq17}
		|\Phi\rangle_{5} &=&\frac{r^{3}}{2}(|+\rangle_{e_{1}}|+\rangle_{e_{2}}|+\rangle_{e_{3}}+|+\rangle_{e_{1}}|+\rangle_{e_{2}}|-\rangle_{e_{3}}\nonumber\\
		&&+|+\rangle_{e_{1}}|-\rangle_{e_{2}}|-\rangle_{e_{3}}+|-\rangle_{e_{1}}|-\rangle_{e_{2}}|-\rangle_{e_{3}}).
	\end{eqnarray}
	For the response of the other detection D$_{q}$ $(q=2,3,4)$, the three-qubit KLM state in Eq. (\ref{eq17}) can be achieved by applying various classical feedforward operations to the output related quantum state, as indicated in Table \ref{Table2}.  Therefore, the success probability  of generating the three-qubit KLM state is $p_{3}=|r^{3}|^{2}$. Notably, the crucial condition for the successful generation of the KLM state is the simultaneous arrival of photon pulses from both the lower and upper arms at the BS.
	
	\begin{table*}
		\centering
		\caption{The measurement results of the ancillary photon state, and the corresponding three-emitter state. The feedforward	operations are given by $I^{e_{3}}=I^{e_{1},e_{2}}$ and $\sigma_{z}^{e_{3}}=\sigma_{z}^{e_{1},e_{2}}$.} \label{Table2}
		\setlength{\tabcolsep}{11pt}
		\renewcommand\arraystretch{2}
		\begin{tabular} {c c c c}
			\hline
			D$_{q}$ & Ancillary photon state & Three-emitter state & Feedforward operation\\
			\hline
			D$_{1}$ & $|H^{9}\rangle$ &
			$-\frac{1}{2}(|+\rangle_{e_{1}}|+\rangle_{e_{2}}|+\rangle_{e_{3}} + |+\rangle_{e_{1}}|+\rangle_{e_{2}}|-\rangle_{e_{3}}$ \\
			& & $+ |+\rangle_{e_{1}}|-\rangle_{e_{2}}|-\rangle_{e_{3}} + |-\rangle_{e_{1}}|-\rangle_{e_{2}}|-\rangle_{e_{3}})$ & $I^{e_{1}}\otimes I^{e_{2}}\otimes I^{e_{3}}$ \\
			D$_{2}$ & $|V^{9}\rangle$ &  $-\frac{1}{2}(|+\rangle_{e_{1}}|+\rangle_{e_{2}}|+\rangle_{e_{3}}-|+\rangle_{e_{1}}|+\rangle_{e_{2}}|-\rangle_{e_{3}}$\\
			& & $-|+\rangle_{e_{1}}|-\rangle_{e_{2}}|-\rangle_{e_{3}}+|-\rangle_{e_{1}}|-\rangle_{e_{2}}|-\rangle_{e_{3}})$ &
			$\sigma_{z}^{e_{1}}\otimes I^{e_{2}}\otimes\sigma_{z}^{e_{3}}$ \\
			D$_{3}$ & $|H^{10}\rangle$ &  $-\frac{1}{2}(|+\rangle_{e_{1}}|+\rangle_{e_{2}}|+\rangle_{e_{3}}-|+\rangle_{e_{1}}|+\rangle_{e_{2}}|-\rangle_{e_{3}}$ \\
			& & $+|+\rangle_{e_{1}}|-\rangle_{e_{2}}|-\rangle_{e_{3}}-|-\rangle_{e_{1}}|-\rangle_{e_{2}}|-\rangle_{e_{3}})$ & $\sigma_{z}^{e_{1}}\otimes\sigma_{z}^{e_{2}}\otimes\sigma_{z}^{e_{3}}$ \\
			D$_{4}$ & $|V^{10}\rangle$ &  $-\frac{1}{2}(|+\rangle_{e_{1}}|+\rangle_{e_{2}}|+\rangle_{e_{3}}+|+\rangle_{e_{1}}|+\rangle_{e_{2}}|-\rangle_{e_{3}}$ \\
			& & $-|+\rangle_{e_{1}}|-\rangle_{e_{2}}|-\rangle_{e_{3}}-|-\rangle_{e_{1}}|-\rangle_{e_{2}}|-\rangle_{e_{3}})$ & $I^{e_{1}}\otimes\sigma_{z}^{e_{2}}\otimes I^{e_{3}}$ \\
			\hline
		\end{tabular}
	\end{table*}

	\begin{figure*}[htpb]
		\centering
		\begin{center}
			\centering
			\includegraphics[width=1\linewidth]{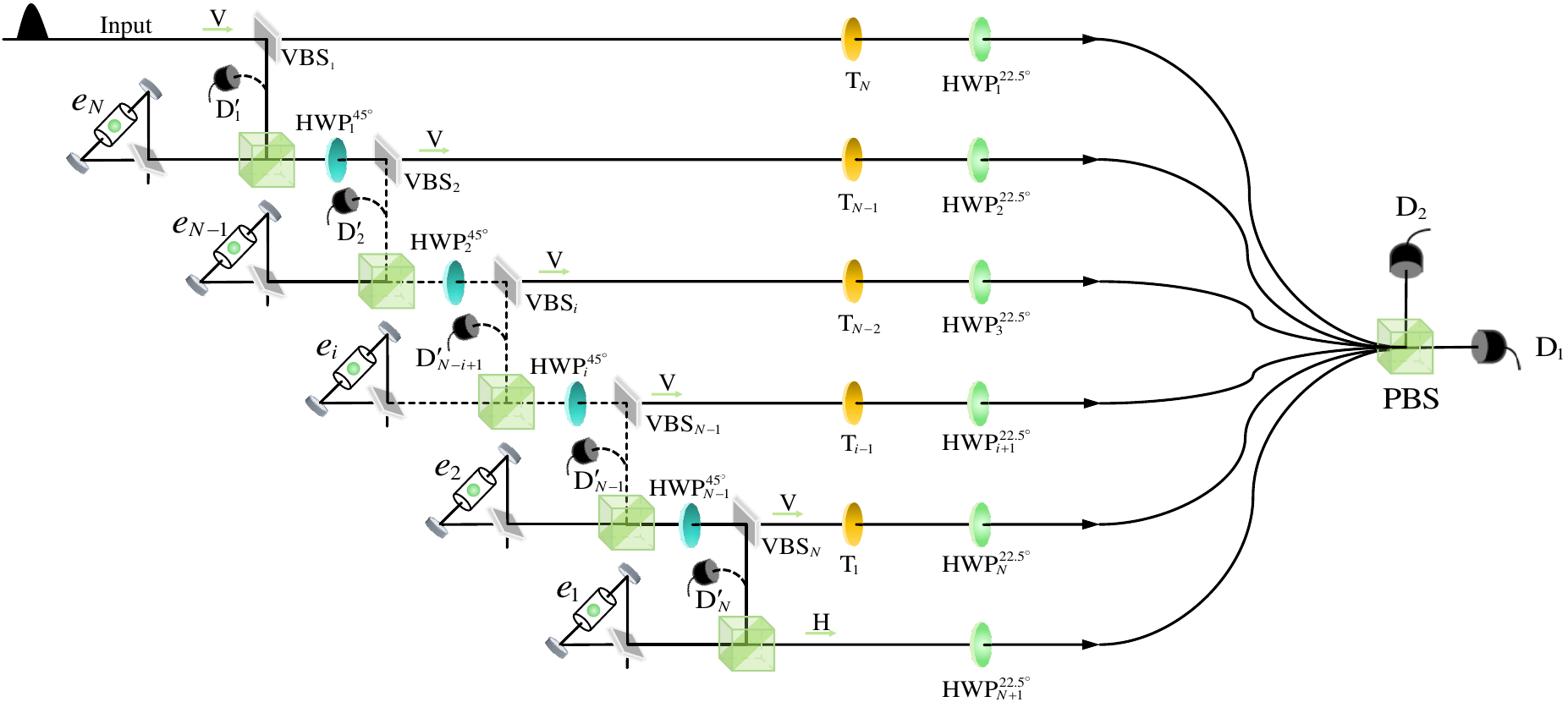}
			\caption{Schematic of the setup for generating $N$-qubit KLM state. }\label{fig4}
		\end{center}
	\end{figure*}
	\subsection{The generation of heralded deterministic $N$-qubit KLM state}\label{sec3.3}
	
	The scheme can be readily extended to the generation of $N$-emitter KLM state, as illustrated in Fig. \ref{fig4}. Initially, the input single-photon state is denoted as $|V\rangle$. The setup consists of  $N-1$ HWPs$^{45^{\circ}}$, $N+1$ HWPs$^{22.5^{\circ}}$ and PBS, $N$ Ts, and $N$ VBSs. Each BS$_{k}(k=1, 2, ..., N)$ with the reflectivity of $(N-k+1)/(N-k+2)$.
	Firstly, after the photon passes through the first VBS$_{1}$,  the photon transmitted by VBS$_{1}$ immediately goes through T$_{N} (j=N)$ with the transmission coefficient $r^{j}$ and HWP$_{1}^{22.5^{\circ}}$. Meanwhile, the photon reflected by VBS$_{1}$ interacts with emitter $e_{N}$ in 1D WG, the state of the photon becomes  $|H\rangle$ without triggering the  detector D$_{1}'$, and it subsequently passes through HWP$_{1}^{45^{\circ}}$ (for converting the photon state into $|V\rangle$) and the second VBS$_{2}$.
	Finally, through the sequence of $N$ steps,
	all output ports are combined at the last PBS to merge $N+1$ spatial modes.
	When detector D$_{1}$ responds, the $N$-emitter KLM state is projected into the standard KLM state,
	\begin{eqnarray} \label{eq18}
		|\Psi_{N}\rangle=\frac{r^{N}}{\sqrt{N+1}}\sum_{j=0}^{N}
		\big(\bigotimes_{1\leqslant a \leqslant j}	|+\rangle_{e_{a}}\big)
		\big(\bigotimes_{j < b \leqslant N}	|-\rangle_{e_{b}}\big),
	\end{eqnarray}
	where $|+\rangle_{e_{a}}$ and $|-\rangle_{e_{b}}(a, b = 1, 2, \dots, N$;$N \geqslant 2$ ) represent the $a$-th and $b$-th emitters are in the states $|+\rangle$ and $|-\rangle$, respectively.
	Conversely, if D$_{2}$ responds, the $N$-emitter KLM state is projected onto the standard $N$-qubit KLM state in Eq. (\ref{eq18}) via a unitary operation $\sigma_{z}^{e_{1}}$ on the emitter $e_{1}$.
	
	\section{Discussion and summary }\label{sec4}
	
	We have developed the heralded schemes to generate  two-qubit and three-qubit KLM states, and  further have popularized to general $N$-qubit KLM state, that leverages the interaction between the auxiliary photon and the emitter 1D WG. Yet, the emitter's decoherence primarily arises from the interaction channels entering non-ideal modes, which can lead to discrepancies between the generated quantum state and the desired KLM states. Additionally, these schemes are influenced by other potential non-ideal factors, such as frequency mismatches, the bandwidth limitations of incident photon pulses, and weak coupling effects between the auxiliary photon and the $N$ emitters. These factors may result in non-ideal interaction scenarios. To mitigate these issues, our schemes incorporate these detectors D$'_{1}$-D$'_{N}$ to monitor the polarization state of the output photon. The detectors identify and discard the photon that undergoes erroneous interaction with the emitter. Specifically, when erroneous interaction occurs, the detector D$'_{i} (i=1,2...,N)$  in Figs. \ref{fig2}-\ref{eq4} detects and removes the incorrect the state of the photon. As a result, the schemes either succeed or fail in a heralded manner, enhancing the generation of  KLM states controllability and reliability. Moreover, to mitigate the effects of photon loss, the single-photon detectors are positioned at the output of generation process of the KLM states.  Therefore, our designs effectively address challenges such as decoherence, non-ideal interactions, and photon loss, making the generation of the KLM states more predictable and reliable.
	
	In the proposed schemes, the core device  for generating the KLM states, as shown in Fig. \ref{fig1}(b), has the success probability that can be calculated using the formula $p_{h}=|\langle\psi|\psi_{r}\rangle|$. Using the relationship $|\psi_{r\rangle}=r|\psi\rangle$, we can derive its success probability  as $p_{h}=|r|^{2}$. Assuming perfect linear optical elements with the success probabilities 100\%, the success probability $p_{h}$ primarily depends on the quality of the waveguide system. Since the device operates based on the heralded mechanism, the successful probabilities for generating the KLM states are influenced by the accuracy of the interaction between the emitter and the photon. The advantage of the heralded mechanism is that errors do not affect the fidelities of these schemes.
	
	\begin{figure}[htpb]
		\centering
		\begin{center}
			\centering
			\includegraphics[width=1\linewidth]{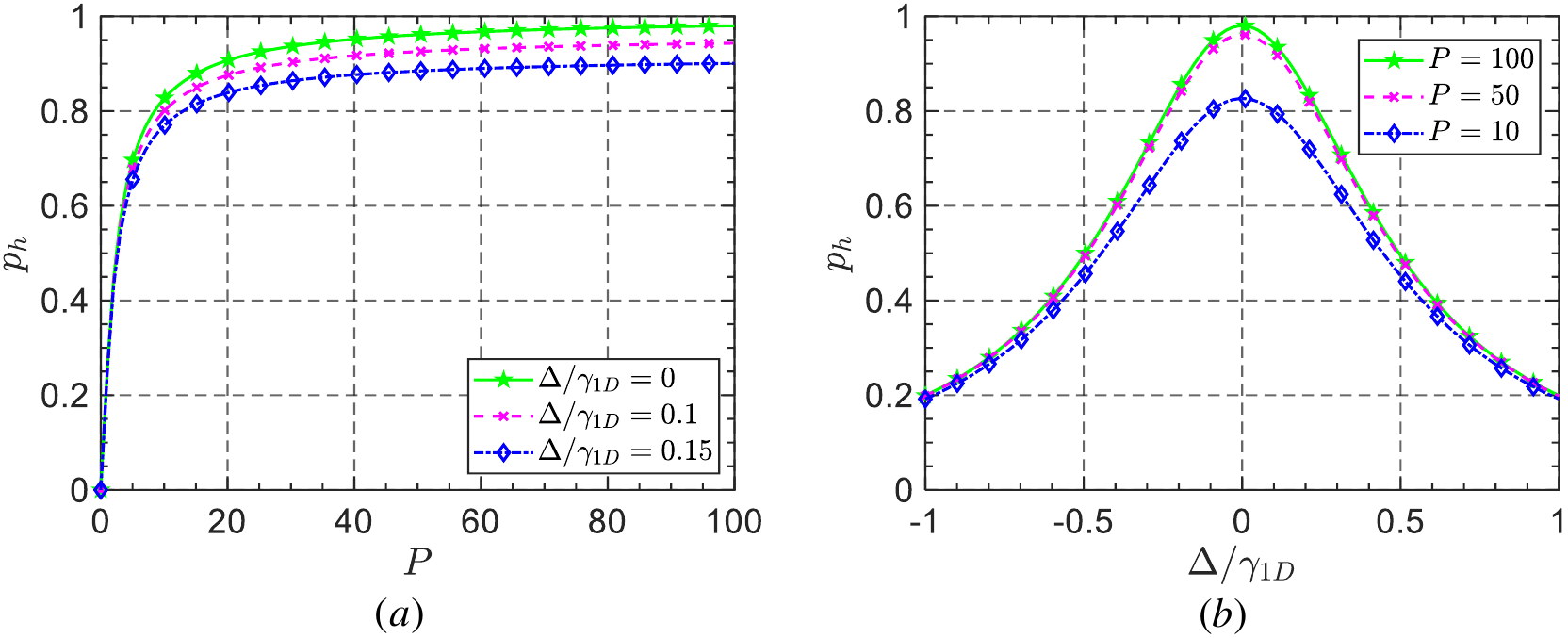}
			\caption{(a) The success probability $p_{h}$ of the heralded device vs the Purcell factor $P$ with fixing the detuning parameter $ \Delta/\gamma_{_{1D}}=0 $ (green solid line), $ \Delta/\gamma_{_{1D}}=0.1 $ (magenta dashed line), and $ \Delta/\gamma_{_{1D}}=0.15 $ (blue dashed-dotted line), respectively. (b)  The success probability $p_{h}$  vs the detuning parameter $ \Delta/\gamma_{_{1D}} $ with fixing the $P = 100$ (green solid line), $P = 50$ (magenta dashed line), and $P = 10$ (blue dashed-dotted line), respectively.}\label{fig5}
		\end{center}
	\end{figure}
	
\begin{figure}[htpb]
		\centering
		\begin{center}
			\centering
			\includegraphics[width=1\linewidth]{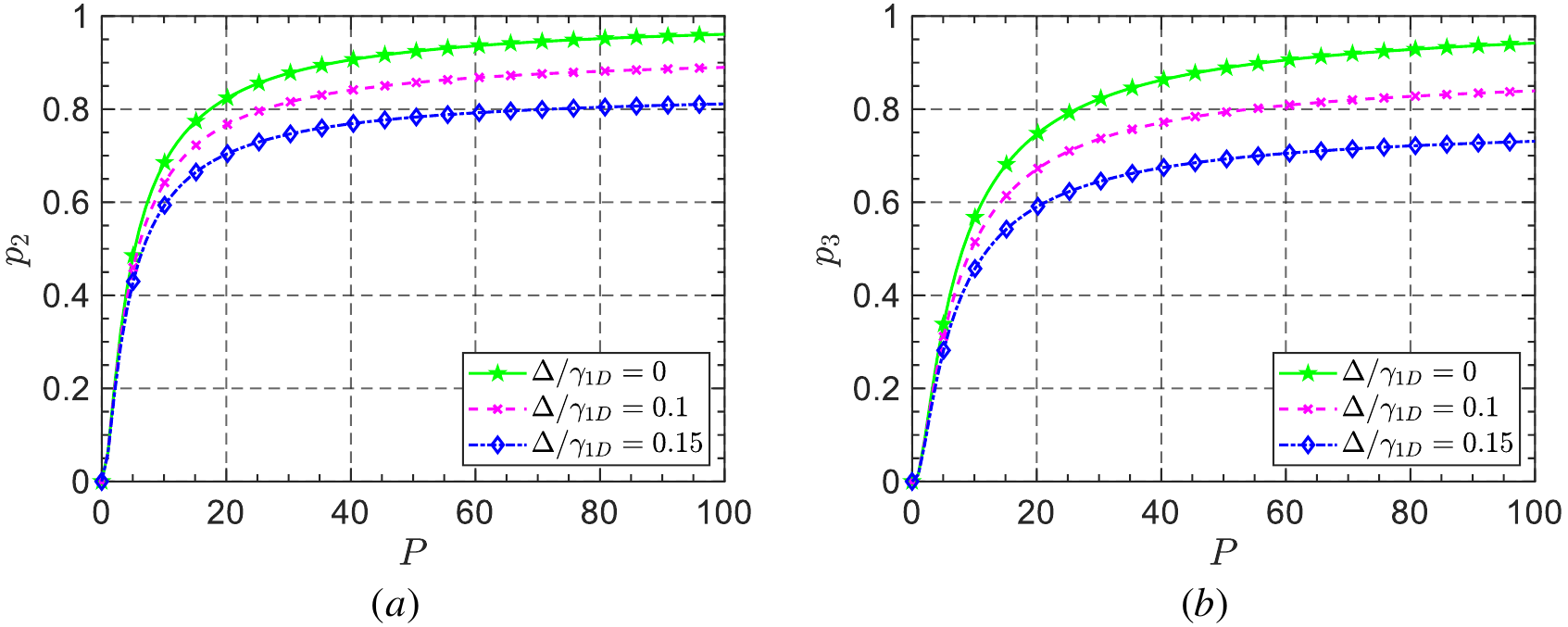}
			\caption{(a) The success probability $p_{2}$ of generating two-qubit KLM state for two emitters (b) the success probability $p_{3}$ of generating three-qubit KLM state for three emitters vs the Purcell factor $P$ with fixing the detuning parameter $ \Delta/\gamma_{_{1D}}=0 $ (green solid line), $ \Delta/\gamma_{_{1D}}=0.1 $ (magenta dashed line), and $ \Delta/\gamma_{_{1D}}=0.15 $ (blue dashed-dotted line), respectively.
			}\label{fig6}
		\end{center}
	\end{figure}

\begin{figure}[htpb]
		\centering
		\begin{center}
			\centering
			\includegraphics[width=1\linewidth]{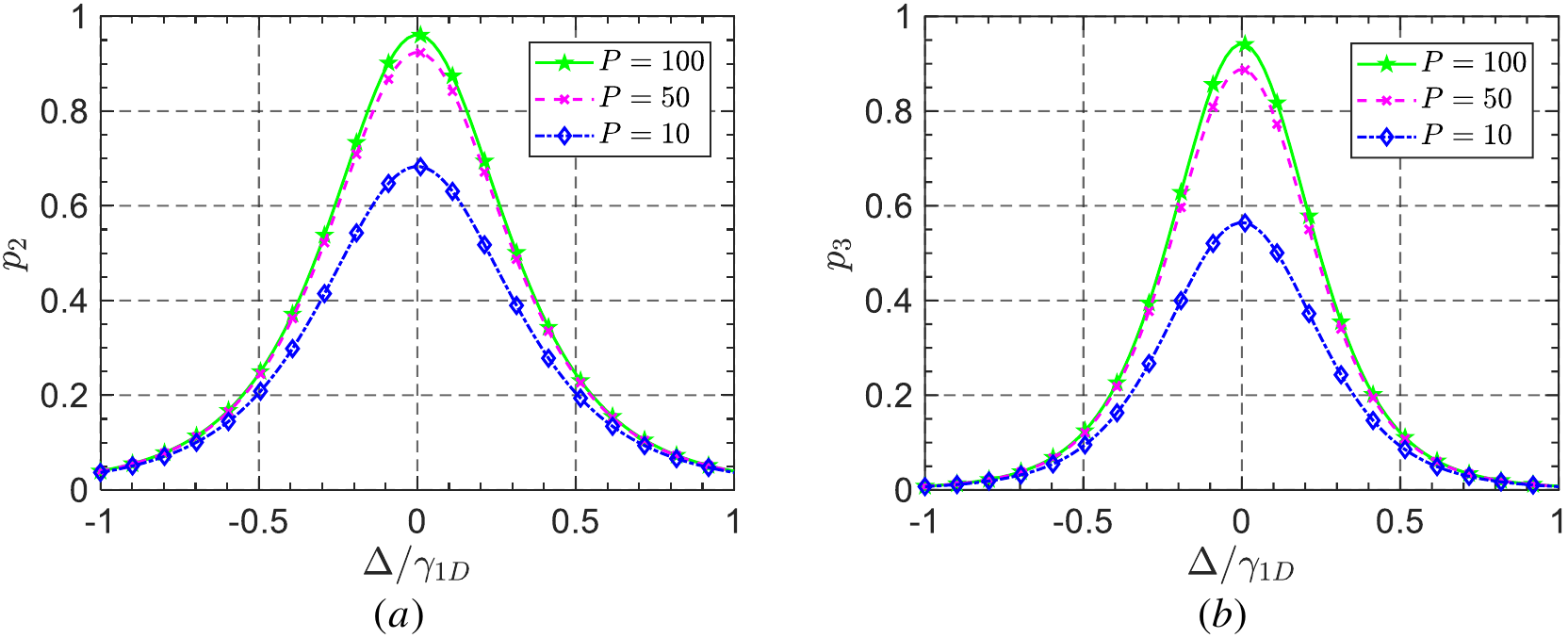}
			\caption{(a) The success probability $p_{2}$ of generating two-qubit KLM state for two emitters (b) the success probability $p_{3}$ of generating three-qubit KLM state for three emitters vs the detuning parameter $ \Delta/\gamma_{_{1D}} $ with fixing the $P = 100$ (green solid line), $P = 50$ (magenta dashed line), and $P = 10$ (blue dashed-dotted line), respectively.}\label{fig7}
		\end{center}
	\end{figure}

	As shown in Fig. \ref{fig5}, we calculate the success probability $p_{h}$ for the heralded Z gate. The results in Fig. \ref{fig5}(a) clearly demonstrate that $p_{h}$ increases significantly with the Purcell factor $P$. Specifically, in the case of  $P\geq50$ and  $|\Delta/\gamma_{_{1D}}| \leqslant 0.13$, the $p_{h}$ surpasses 90\%, suggesting that these schemes are feasible for practical applications. For example, with $P=100$ and $\Delta/\gamma_{_{1D}} = 0.1$, the $p_{h}$ of the 1D WG system reaches 94.33\%. Additionally,  the success probability $p_{2}$ of generating two-qubit KLM state and the success probability $p_{3}$ of generating three-qubit KLM state are analyzed as  the Purcell factor $P$ and the frequency detuning $\Delta/\gamma_{_{1D}}$, as presented in Figs. \ref{fig6} and \ref{fig7}, respectively. For instance, when $P=100$ and
	$\Delta/\gamma_{_{1D}} = 0$, $p_{2}$  is 96.10\% and $p_{3}$  is 94.20\%. However, as the frequency detuning increases to
	$\Delta/\gamma_{_{1D}} = 0.15$, $p_{2}$  drops to 81.15\% and $p_{3}$  falls to 73.10\%. Moreover, for the fixed $\Delta/\gamma_{_{1D}}$, an increase of the $P$ significantly enhances $p_{2}$ and $p_{3}$. For example, when $P=10$ and $\Delta/\gamma_{_{1D}} = 0$, $p_{2}$ and $p_{3}$ are 68.26\% and 56.39\%, respectively. In contrast, when the
	$P$ is increased to $100$ with $\Delta/\gamma_{_{1D}}=0$, $p_{2}$ and $p_{3}$ rise to 96.02\% and 94.10\%, respectively.
	In a word, the study demonstrates that, decreasing $\Delta/\gamma_{_{1D}}$ and magnifying $P$ can
	substantially improve $p_{2}$ and $p_{3}$. Furthermore, as the number of entangled qubits increases,  two factors have a greater impact on  the success probability $p_{k} (k=2,3,...,N)$.
	
	\begin{figure}[htpb]
		\centering
		\begin{center}
			\centering
			\includegraphics[width=1\linewidth]{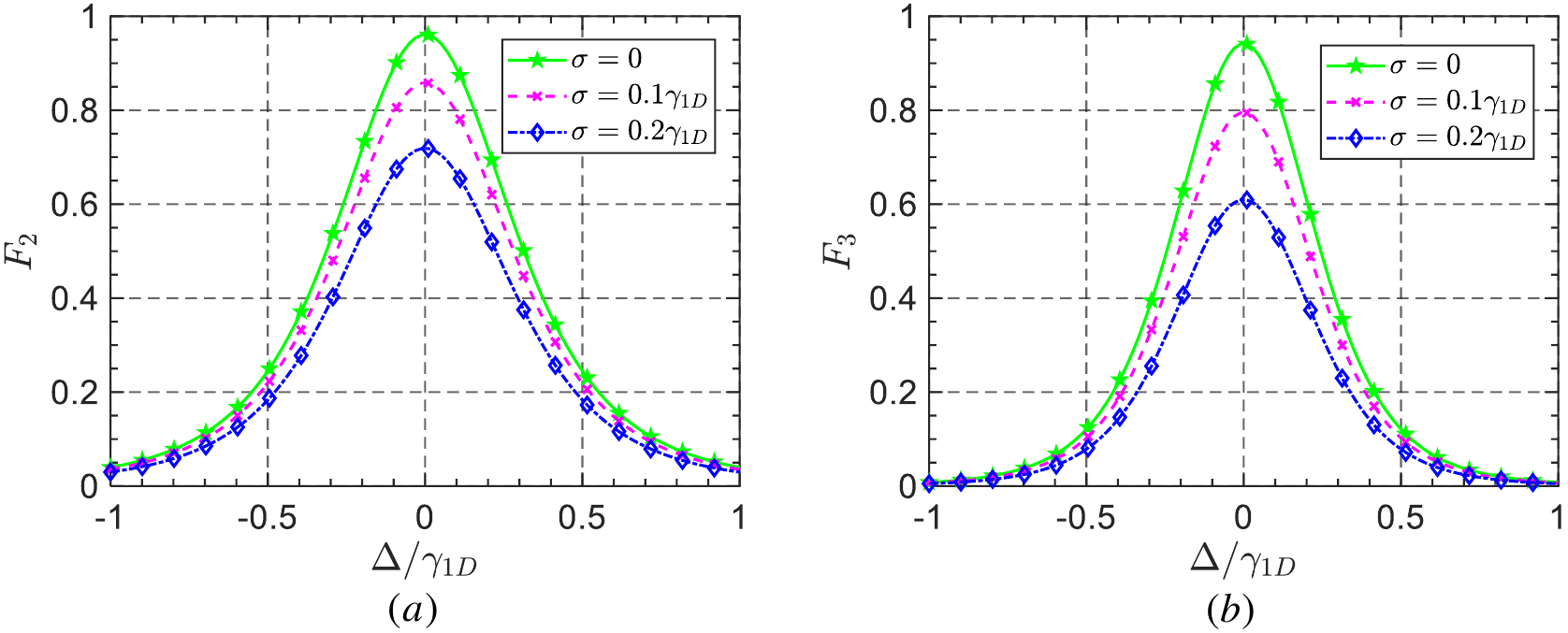}
			\caption{(a) The fidelity $F_{2}$ of generating two-qubit KLM state for two emitters (b) the fidelity $F_{3}$ of generating three-qubit KLM state for three emitters vs the detuning parameter $ \Delta/\gamma_{_{1D}} $ with fixing $P = 100$ and the inhomogeneous detuning $\sigma = 0$ (green solid line), $\sigma = 0.1\gamma_{_{1D}}$ (magenta dashed line), and $\sigma = 0.2\gamma_{_{1D}}$ (blue dashed-dotted line), respectively.}\label{fig8}
		\end{center}
	\end{figure}
	
	In practical scenarios, optical transitions in the emitter are commonly affected by inhomogeneous broadening \cite{mittelstadt2021terahertz,basso2019spectral}, which can degrade the fidelities of our proposed schemes. 
	To address this issue, we assume that the inhomogeneous broadening follows a Gaussian distribution, represented by the probability density function $\rho(\delta)=\dfrac{1}{\sqrt{2\pi\sigma^{2}}}\exp(-\dfrac{\delta^{2}}{2\sigma^{2}})$, where $\delta$ denotes the inhomogeneous detuning, and $2\sigma$ corresponds to the full width at half maximum. Figure \ref{fig8} demonstrates the effect of inhomogeneous broadening on the KLM state generation schemes under three conditions: Purcell factor $P = 100$, with $\sigma$ values of $0$, $0.1\gamma_{_{1D}}$, and $0.2\gamma_{_{1D}}$. As shown in Fig. \ref{fig8}, for a fixed frequency detuning, an increase in inhomogeneous broadening (i.e., with increasing $\sigma$) leads to a decrease in two fidelities. However, as the frequency detuning increases, the influence of inhomogeneous broadening on two fidelities becomes less pronounced. Therefore, to maintain high fidelities in schemes, it is essential to effectively suppress the inhomogeneous broadening.

	In summary, we have proposed three heralded schemes for the generation of two-qubit, three-qubit, and $N$-qubit KLM states, leveraging the interaction between the single photon and  solid-state emitters. In these approaches, imperfect interaction events caused by system defects,  such as frequency mismatches, the bandwidth limitations of incident photon pulses, and weak coupling effects, can be identified and discarded through the  heralded mechanism, leading to exhibiting high fidelities. With ongoing advancements in waveguide systems, our proposals not only contribute to a deeper understanding and application of quantum entanglement but also open new avenues for future research in quantum computing.

	\begin{acknowledgments}
		This work was supported in part by the Natural Science Foundation of China under Contract 61901420;	
		in part by Fundamental Research Program of Shanxi Province under Contract 20230302121116.
	\end{acknowledgments}
	
	\section*{Disclosures}
	The authors declare no conflicts of interest.
	
	\section*{Data Availability Statement}
	Data underlying the results presented in this paper are not publicly available at this time but may be obtained from the authors upon reasonable request.

	
	\bibliography{PRAreferences}
	
\end{document}